# Integrated silicon optomechanical transducers and their application in atomic force microscopy

Jie Zou[1, 2], Marcelo Davanco[1, 3], Yuxiang Liu[1, 2, 4], Thomas Michels[1, 2], Kartik Srinivasan[1], and Vladimir Aksyuk[1]

[1] Center for Nanoscale Science and Technology, National Institute of Standards and Technology, Gaithersburg, Maryland 20899, USA

[2] Maryland Nanocenter, University of Maryland, College Park, MD 20742, USA

[3] Department of Applied Physics, California Institute of Technology, Pasadena, CA 91125, USA

[4] Department of Mechanical Engineering, Worcester Polytechnic Institute, Worcester, MA 01609, USA

## 1.1 Introduction

Physics experiments measuring unknown quantities by transducing them to mechanical motion have a long and distinguished history[1]. The advent of micro- and now nano-fabricated mechanical transducers has continued this trend over the last two decades, where miniaturization enabled better coupling and measurement of microscopic physical phenomena, from electron and nuclear spins[2,3] to individual vortices in superconductors[4], from quantum vacuum fluctuations in nanoscale optical cavities[5,6] to shapes and masses of individual molecules[7]. While micromechanical rotation and acceleration sensors are now ubiquitous in both cars and cell phones, in the physics laboratory nanoscale mechanical devices have now been cooled to their quantum mechanical ground state[8,9] and continue to enable measurements with an unprecedented degree of precision and control. Far from being confined to a narrow set of unique applications, micromechanical measurement is at the center of the Atomic Force Microscope (AFM)[10,11], with its wide variety of operation modes and applications in physics, biology, and industrial inspection[12].

Fabrication techniques for nanobeams and more complicated mechanical transducers with one or more critical dimensions below 100 nm are now well established. Due to their small size and low mass, such devices can achieve a unique combination of high speed, high sensitivity, and local coupling to nanoscale systems and phenomena. Mechanical resonance frequencies above 100 MHz can be achieved without sacrificing mechanical compliance and force sensitivity[2-7]. The ultimate limit on the performance of such transducers is imposed by the fundamental thermodynamic mechanical force noise in accordance with the fluctuation dissipation theorem. It is therefore critical to minimize mechanical dissipation in the transducer. With proper materials and fabrication techniques, a very high mechanical quality factor ($Q_m$) in the range of $10^4$ to $10^6$ can be achieved in vacuum environment. In combination with the very low mass and stiffness, phenomenally low intrinsic noise can be achieved, even at room temperature. Although when operated in ambient, the air damping significantly reduces the $Q_m$, the absolute value of the damping coefficient and the corresponding force sensitivity can scale favorably compared to larger mechanical transducers.

However, one of the most significant roadblocks to realizing the full potential of nanomechanical sensing is the readout of the motion of a small transducer with high sensitivity, high bandwidth, and without excess power dissipation. Electrical means of motion readout[13], such as capacitive, magnetomotive, piezoresistive, and piezoelectric have been successfully employed, but all of them suffer from various combinations of poor scaling with reduced size, power dissipation limitations, magnetic field and materials requirements, and all introduce thermal Johnson noise into the readout signal. On the other hand, optical techniques do not suffer from thermal noise, do not in principle need to dissipate any power at the transducer, and have an enormous measurement bandwidth. By introducing optical cavities, light is trapped and made to interact longer with the mechanical transducer, allowing matching the bandwidth of optical readout to the mechanical transduction bandwidth[14-16]. The reduction in the readout bandwidth is traded for a drastic increase in the readout gain.

Effective coupling of a microscopic transducer to an external, bulk optical cavity is technically challenging due to alignment and drift. Moreover, the need for a high reflectivity coating on the transducer may introduce mechanical dissipation and reduce the sensitivity[14]. Furthermore, because of the diffraction limit, in such a system the light cannot be focused to a spot much smaller than the optical wavelength. When the transducer (cantilever, nanobeam, etc.) size is reduced below the wavelength of light, the coupling and the sensitivity deteriorate dramatically. These challenges can be overcome by integrating a nanomechanical transducer in the near-field of a nanophotonic cavity microfabricated together on the same chip[17-19]. The cavity can be self-aligned to the nanoscale cantilever with an accuracy of a few nanometers, and the cantilever can interact strongly with the optical field in the cavity without mechanical contact. It is also unnecessary to introduce any coating, so that the mechanical quality factor is only limited by intrinsic dissipation of the material. The optical quality factor of the cavity can be maintained as well. This fully integrated, stable and practical optomechanical device can be fiber connectorized. The end result is the implementation of a nanoscale transducer with GHz bandwidth and precision near the standard quantum limit, all while employing only microwatts of optical power.

Such combined sensors can operate at the fundamental thermomechanical noise floor not only near the mechanical resonance frequency, but over a broad frequency range from near DC to several times the mechanical resonance frequency[19]. Furthermore, the dynamics of the mechanical transducer can be modified and tailored to meet specific needs by using optical forces, or another auxiliary actuation mechanism such as an integrated electrostatic actuator. Both a classical feedback control scheme with virtually no excess noise injected into the transducer, and a quantum control scheme through optical forces, whereby the transducer acts directly on the cavity optical field, can be implemented. Such schemes can be used to tune the mechanical resonant frequency over a wide range, implement a regenerative oscillator, or cold-dampen a high mechanical $Q_m$, low-noise transducer and produce a flat transfer function for operation over a broad frequency range.

This chapter is arranged as follows: first we will discuss the device design, the transduction scheme, and the measurement result of a typical device. Then we will discuss numerical simulations in detail. In Sec. 1.4, we show the progress towards the application of such devices as AFM probes. Finally we conclude with a summary and outlook.

## 1.2 Optomechanical Transduction and Device Fabrication
### 1.2.1 Design and Transduction Scheme

Typically, optical readout approaches for measuring cantilever motion, such as beam deflection[20] and laser interferometry[14], rely upon free-space optics and make measurements in the far-field. In contrast, the integrated sensor approach (Fig. 1a) utilizes near-field interaction as a probe of nanocantilever motion. One suitable optical readout tool is a silicon microdisk resonator, which is a device that supports 'whispering gallery' optical modes. These modes circulate around the microdisk edge and have evanescent tails that extend out into the surrounding air cladding. Introducing a cantilever into this evanescent region induces a shift in the optical mode frequencies $\omega_{opt}$, with the amount of shift depending on the specific location of the cantilever with respect to the disk. Thus, as the cantilever vibrates, the resonant frequency of a given optical mode varies, and this in turn can be mapped to a varying optical intensity in a number of ways. One straightforward approach[15] is to use a laser that is tuned to the shoulder of the cavity mode optical resonance, as shown in Fig. 1b. Alternately, measurements of phase on resonance with the cavity mode can be used.

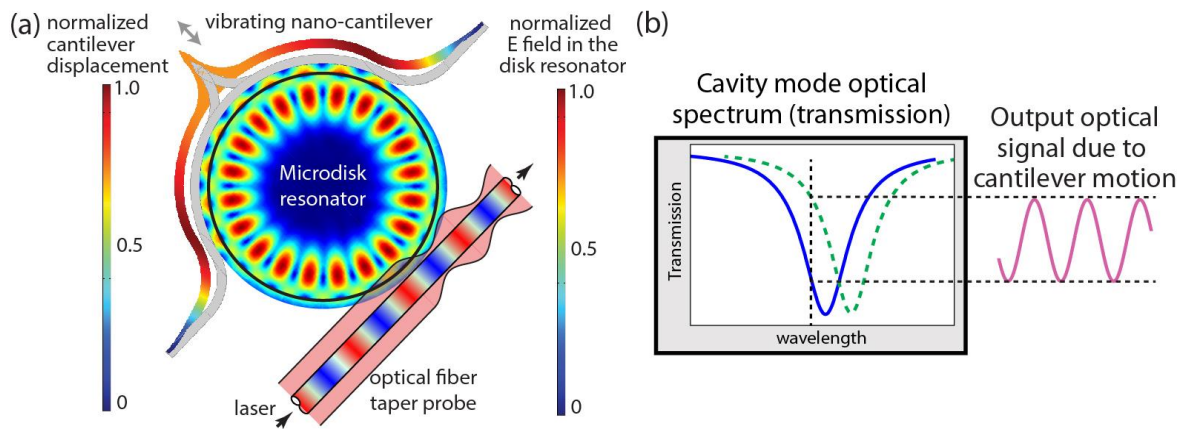

Fig. 1. (a) Working principle of the disk-cantilever optomechanical sensor. The cantilever's equilibrium position is depicted in grey. The colored cantilever shows the FEM-calculated deformed shape (with an exaggerated amplitude) of the first order, in-plane, even-symmetry mechanical mode, for a system with disk diameter of 2.5 μm, cantilever width of 125 nm, and cantilever thickness of 260 nm. The color map in the microdisk resonator represents the absolute value of the FEM-calculated electric field amplitude of the relevant optical mode. The left scale bar is for the cantilever displacement, the right one for the electric field amplitude in the microdisk, and the electric field in the fiber probe is not in scale. (b) Using a tunable laser whose wavelength is aligned to the shoulder of an optical mode enables fluctuations in the cavity mode optical frequency due to cantilever motion to be mapped to an intensity-modulated optical signal. Adapted with permission from ref. 18.

This approach is characterized by several parameters that determine the optical readout sensitivity. The first is the optomechanical coupling parameter $g_{OM} = d\omega_{opt}/dx$, which represents the change in the microdisk mode's optical frequency $\omega_{opt}$ per unit change in the disk-cantilever separation $x$. This parameter thus determines the amount of frequency shift induced in the optical mode when the cantilever vibrates. Next, the optical quality factor ($Q_{opt}$) of the mode is important. Qualitatively, one expects the minimum detectable frequency shift to be set by (a fraction of) the cavity mode optical linewidth. More precisely, the cavity mode lineshape determines the conversion between the frequency modulation created by the cantilever motion and the intensity modulation produced when the input laser is tuned to the shoulder of the microdisk optical mode (Fig. 1b). Other important parameters include the out-coupled optical power from the microdisk-cantilever sensor, the noise equivalent power of the photodetector used, the power and wavelength noise of the laser source, and the stability of the optical transducer with respect to thermal drift.

Sensor geometries should be chosen to optimize parameters such as $g_{OM}$ and $Q_{opt}$. Using a semicircular cantilever shape increases the interaction length between the microdisk optical mode and the cantilever's mechanical mode with respect to what can be achieved with a straight cantilever. This is verified by finite element method simulations[18] (see Sec. 1.3 for more details), reproduced in Fig. 2a, where $g_{OM}$ for the curved cantilever geometry is seen to exceed the value calculated for a straight cantilever by about one order of magnitude. Typical values for such a system are $g_{OM}/2\pi \approx 1$ GHz/nm to 10 GHz/nm at gaps $G \approx 100$ nm. In addition, the semicircular cantilever shape largely preserves the low optical loss possible in Si microdisks[21], so that $Q_{opt} \geq 10^4$ can be readily achieved. We note that the requirement of high $Q_{opt}$ sets a limit on the cantilever width we can use. As the width increases above $\approx 300$ nm, the cantilever changes from having a perturbative effect on the microdisk mode to having a much stronger influence. In particular, once the cantilever width becomes large enough to support optical modes, optical quality factors of the microdisk deteriorate dramatically.

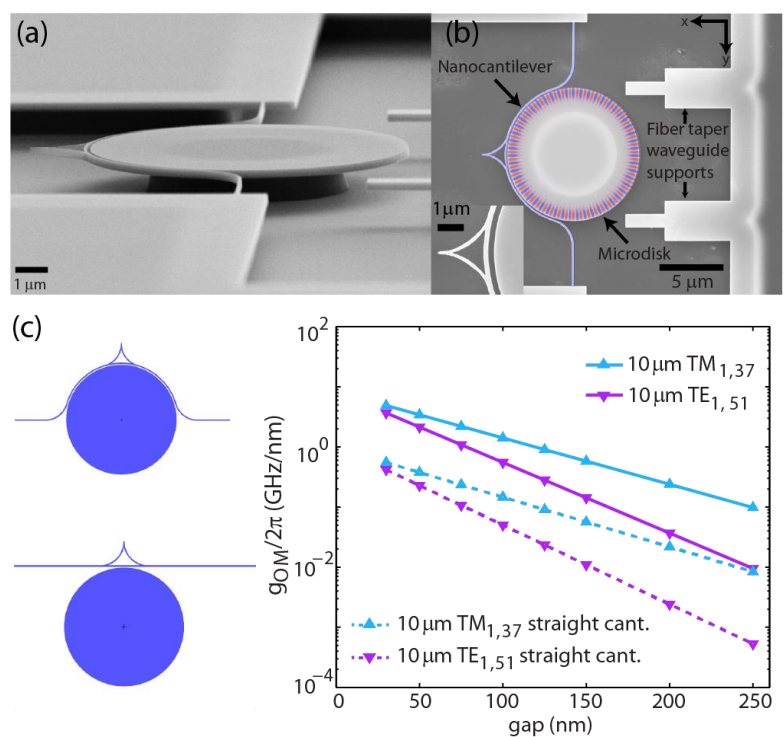

Fig. 2. Scanning electron micrograph of a fabricated device ($D = 10$ μm) from (a) a tilted angle and (b) the top view. The inset: a close-up of the gap and width of the cantilever near the tip. (c) Calculated optomechanical coupling parameter $g_{OM}$ as a function of disk-cantilever gap, for different optical modes ($TM_{1,37}$, $TE_{1,51}$) and the two different cantilever shapes (curved and straight) shown in the diagram to the left. Panels (a, b) are adapted with permission from ref. 17. Copyright (2011) American Chemical Society. Panel (c) is adapted from ref. 18.

### 1.2.2 Fabrication

In experiments focused on establishing the basic optomechanical transduction mechanism, a simplified fabrication process was used, consisting of electron beam lithography (needed to establish $\approx 100$ nm gaps and cantilever widths), inductively-coupled plasma reactive ion etching of the silicon layer, hydrofluoric acid (HF) wet etching of the sacrificial oxide layer, and a liquid $CO_2$ critical point drying. These devices, shown in Fig. 2a, are suitable for determining the thermal noise spectrum of different cantilever geometries, and the sensitivity with which the cantilever motion can be optically read out by the microdisk resonator. As these devices do not have access waveguides fabricated on-chip, optical fiber taper waveguides are used to couple light into and out of the devices. Experiments utilizing the device in a scanning probe configuration require a more extensive fabrication process, which will be elaborated in Sec. 1.4.

### 1.2.3 Detection Setup

A detection setup for characterizing these devices is shown in Fig. 3a. Light from a tunable laser is coupled into the device using, for example, an optical fiber taper waveguide. The fiber taper waveguide was fabricated by heating and stretching the optical fiber down to ≈1 μm in diameter. A local indentation ("dimple") with ≈10 μm radius of curvature is formed within the thinnest region of the fiber[22], allowing for selective probing of devices within two-dimensional arrays. A swept-wavelength laser with a wavelength range of 1520 nm to 1630 nm was used as the light source, and was sent into a polarization controller before going into the fiber taper waveguide, allowing for polarization adjustment to maximize the coupling depth of the desired optical mode before recording data. Light circulates within the microdisk resonator hundreds or thousands of times (depending on the cavity's finesse) before exiting back through the same fiber taper waveguide. The output of the fiber is detected with a low bandwidth photodetector, and the transmission spectrum of the device is recorded, revealing the spectral location and spectral width of the cavity's optical modes. As described above, motion of the cantilever results in a frequency modulation of the optical cavity modes, which can be translated into an intensity modulation by probing these modes on the side of their resonance minima. The information obtained from the transmission spectra is thus used to determine the laser wavelength for optimal transduction sensitivity. The output signal exiting the device, which now carries the imprint of the mechanical motion as an intensity fluctuation, is detected on a second, higher bandwidth photodetector before being sent to an electronic spectrum analyzer to reveal the spectrum of mechanical modes.

Figure 3b displays the thermomechanical noise spectrum of one typical device with a microdisk diameter $D = 10$ μm. The thermal motion of the first 6 in-plane mechanical modes (see next section for the simulated mode shapes) is observed and the results are summarized in Table 1. The FEM simulation was used to calculate the theoretical mechanical resonance frequency $f_m$, and the effective mass $m_{eff}$, based on the measured dimensions. $Q_{opt}$ is obtained from the experimental optical spectra. The experimental $f_m$ and $Q_m$ were obtained by fitting Lorentzian functions to the experimentally acquired mechanical spectra around the resonance peaks. The experimental and simulation results are in excellent agreement.

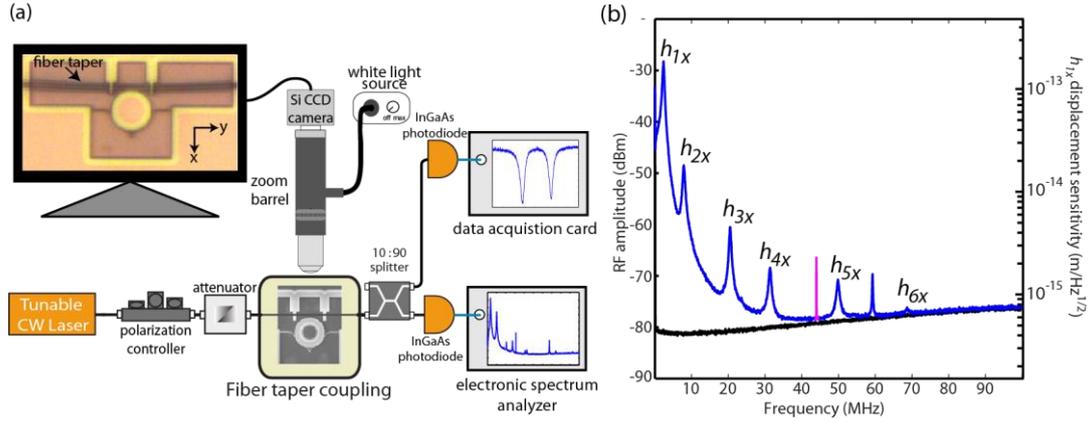

Fig. 3. (a) Typical experimental setup. Light from a tunable laser is sent through a polarization controller and variable optical attenuator before being coupled into the device, in this case using an optical fiber taper waveguide. The light exiting the device is coupled back out into the same optical fiber taper waveguide, and then split into two channels with a fiber coupler. The first channel is used with a low bandwidth photodetector, and is used to measure the optical cavity's transmission spectrum in situations in which the laser wavelength is swept. The second channel is used with a high bandwidth photodetector, and is used in situations in which the laser wavelength is fixed on the shoulder of the optical cavity mode. The output of the high bandwidth photodetector is fed to an electronic spectrum analyzer to resolve the cantilever mechanical modes. (b) The thermomechanical noise spectrum is shown for a typical disk-cantilever device ($D$ = 10 μm, $w$ = 65 nm, $G$ =100 nm). Six in-plane mechanical modes are identified. The red spike corresponds to an electrical driving signal that is applied for calibration purposes. The black curve represents the measured detector noise. Reprinted with permission from ref. 17. Copyright (2011) American Chemical Society.

| mode | Cal. $k$ (N/m) | Cal. $m_{eff}$ (pg) | Exp. $f_m$ (MHz) | Cal. $f_m$ (MHz) | Exp. $Q_m$ |
|---|---|---|---|---|---|
| $h_{1x}$ | 0.14 | 0.73 | 2.23 | 2.35 | 4.9 |
| $h_{2x}$ | 1.41 | 0.58 | 7.82 | 7.89 | 13.1 |
| $h_{3x}$ | 5.72 | 0.35 | 20.37 | 20.51 | 38.5 |
| $h_{4x}$ | 12.43 | 0.32 | 31.17 | 31.36 | 44.4 |
| $h_{5x}$ | 41.95 | 0.44 | 49.36 | 49.86 | 61.2 |
| $h_{6x}$ | 74.05 | 0.40 | 68.13 | 68.71 | 91.0 |

Table 1. Experimentally measured and calculated properties for in-plane mechanical modes of a typical disk-cantilever device ($D$ = 10 μm, $w$ = 65 nm, $G$ =100 nm). Adapted with permission from ref. 17. Copyright (2011) American Chemical Society.

## 1.3 Numerical Simulation

Calculation of the important parameters, such as $\omega_{opt}$, $Q_{opt}$, $f_m$, and $g_{OM}$, enables optimization of the microdisk-cantilever design. Optical and mechanical resonances of an optomechanical microcavity can be calculated by solving the appropriate eigenvalue equations, respectively, for the electromagnetic field and mechanical displacement. Then one can obtain the optomechanical coupling $g_{OM}$ from the solutions of the electromagnetic field and mechanical displacement. In this section, we briefly describe the numerical methods and their application to our geometries.

### 1.3.1 Optical resonances

For optical resonances, the eigenvalue equation is

$$\nabla \times \nabla \times \mathbf{E} = \varepsilon(\mathbf{r})\left(\frac{\omega}{c}\right)^2 \mathbf{E}, \qquad (1)$$

assuming a time-harmonic electric field $\mathbf{E} = \mathbf{E}(\mathbf{r})\exp(i\omega t)$. In Eq. 1, $\varepsilon(\mathbf{r})$ is the permittivity distribution, $c$ is the speed of light and $\omega$ is the resonance frequency. The permittivity distribution $\varepsilon(\mathbf{r})$ is known and essentially defines the optical resonator geometry. Solving Eq. 1 produces a discrete set of solutions with frequencies $\omega_i$ and corresponding field distributions (eigenvectors) $\mathbf{E}_i$, which are the optical cavity resonances. While in some simplified geometries Eq. 1 can be solved analytically, generally a numerical partial differential equation solving method, such as Finite Difference or Finite Elements[23] is employed. Finite difference and finite elements are the most popular – but not the only – methods for PDE solving, and are widely available in both free and commercial implementations.

An important aspect of optical cavities is that in general the cavities are "open", meaning that the optical field extends past the physical boundaries of the cavity. In fact the field extends infinitely into the region surrounding the geometry. Therefore, ideally Eq. 1 must be solved over the entire (infinite) space. A variety of methods have been devised to reduce the solution domain to practically realizable sizes, generally involving the enforcement of special field conditions at the boundaries of the (finite) computational window, that locally approximate the form of the real solution. Currently, however, the most popular method for limiting the size of the computational window is the Perfectly Matched Layer (PML) method[24]. Essentially, it consists of surrounding the computational window with a thin layer of an artificial material that is both absorbing and optically matched to the medium external to the optical cavity. Because the PML is optically matched to the medium it surrounds, waves leaking from the cavity and impinging upon it are not reflected, and are absorbed once inside the PML.

Power leakage from an optical cavity is characterized by the cavity quality factor $Q_{\text{opt}}$, defined as $Q_{\text{opt}} = \omega_{opt} \cdot \tau_p$, where $\tau_p$ is the cavity photon lifetime. This is essentially a measure of how many resonance cycles are necessary for a photon to leave the cavity[25], or for the energy stored in the cavity to drop to $e^{-2\pi}$. The cavity $Q_{\text{opt}}$ can be obtained from Eq. 1 by solving it with appropriate boundary conditions, which as discussed above, simulate free space. In this case, complex eigenfrequencies $\tilde{\omega}$ are obtained, and (given the time-harmonic form of the electric field) the quality factor can be obtained as $Q_{\text{opt}} = \text{Re}(\tilde{\omega})/2\,\text{Im}(\tilde{\omega})$.

An alternative, very popular technique for calculating optical modes is the Finite Difference Time Domain method[26]. Here, one solves a discretized (finite difference) version of Maxwell's curl equations in time domain, in contrast to solving the frequency-domain eigenvalue problem of Eq. 1. The curl equations for the electric and magnetic fields are mutually coupled, and both E and H are calculated simultaneously on a grid at each time-step. This typically is done using the Yee algorithm (after K. Yee, who pioneered the technique[26, 27]), in which a grid for E and H is defined such that the E components are surrounded by circulating H components and vice-versa.

Resonant modes can be calculated by exciting the cavity with a spectrally broad source and monitoring the field decay in time. These modes are long lived, and appear as peaks in the Fourier transform of the time evolution of the field. Typically, once a resonant mode is identified, subsequent simulations are performed in which the same mode is excited with a spectrally narrow source centered at its frequency, so that the field evolution is given only by the target resonance. As in the frequency-domain case, the computational domain is truncated, and PMLs are typically used to simulate open spaces. The radiation-limited quality factor of the cavity can be obtained either by monitoring the decay of the energy in the cavity and the power loss through the computational window boundaries, all of which can be calculated from the electromagnetic field, or also by fitting the time-evolution of one or more field components at specific points within the cavity.

Simulation results presented in this chapter were obtained using the finite element method.

**1.3.2 Mechanical resonances**

Mechanical resonances are solutions[28, 29] to the eigenvalue equation for the mechanical displacement **u**(**r**)

$$\nabla \cdot \left[ \mathfrak{C} : \frac{(\nabla + \nabla^T)}{2} \mathbf{u} \right] = \rho \omega^2 \mathbf{u}, \tag{2}$$

where ρ is the mass density and $\mathfrak{C}$ is the elasticity tensor, which is of rank four. Both $\mathfrak{C}$ and **u** are spatially variant and effectively define the geometry. For isotropic materials,

$$\frac{1}{\mathfrak{C}} = \begin{bmatrix} 1 & -\nu & -\nu & 0 & 0 & 0 \\ -\nu & 1 & -\nu & 0 & 0 & 0 \\ -\nu & -\nu & 1 & 0 & 0 & 0 \\ 0 & 0 & 0 & 2\cdot(1+\nu) & 0 & 0 \\ 0 & 0 & 0 & 0 & 2\cdot(1+\nu) & 0 \\ 0 & 0 & 0 & 0 & 0 & 2\cdot(1+\nu) \end{bmatrix} \tag{3}$$

and the double-dot product corresponds to a fourth-rank to second-rank tensor product. The eigenvalue equation (Eq. 2) produces a discrete set of frequencies $\omega_m$ and corresponding displacement fields **u**(**r**), which are the mechanical modes of the simulated structure. Equation 2 is typically solved with the finite element method. Using a commercial FEM software, we can calculate the mode shapes [relating to **u**(**r**)] and the mechanical frequencies $f_m$. Figure 4 displays the mode shapes of the first four in-plane modes with the displacement mainly along the *x* direction.

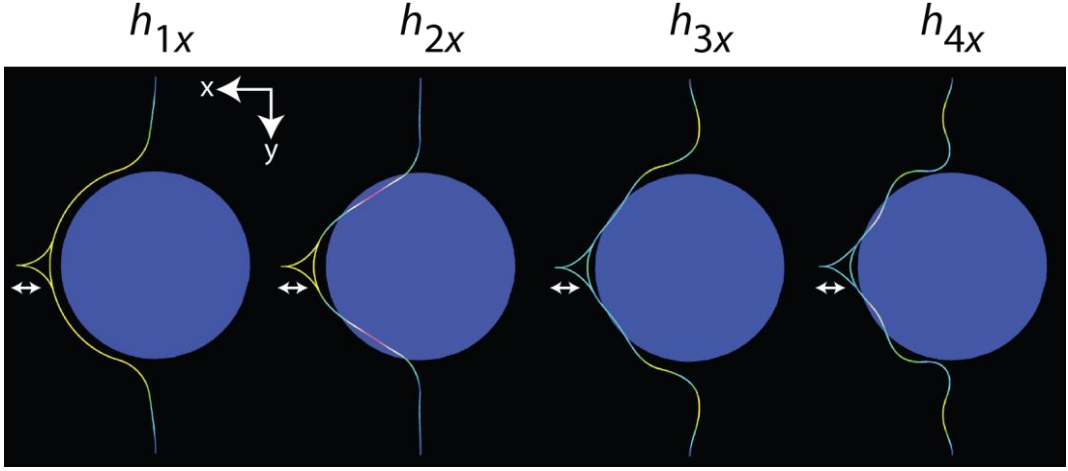

Fig. 4. Numerically calculated mechanical mode shapes (amplitude exaggerated for clarity) with dominant displacement along the x axis for a device with *D* = 10 μm and *w* = 65 nm. Reprinted with permission from ref. 17. Copyright (2011) American Chemical Society.

### 1.3.3 Optomechanical coupling rate

The shift in the frequency $\omega_{opt}$ of a particular optical resonance due to the displacement of the nanostructure boundaries produced by a mechanical resonance at frequency $f_m$ is quantified by the optomechanical coupling $g_{OM} = \partial \omega_{opt}/\partial x = \omega_{opt}/L_{OM}$; here, $x$ is the cavity boundary displacement and $L_{OM}$ is an effective optomechanical interaction length. For the disk-cantilever work presented in this chapter, the following procedure was followed for calculating $g_{OM}$. First, the mechanical modes of the cantilever were simulated to determine the frequency and shape of the mode of interest, which is the in-plane even-symmetry mode. The cantilever was then deformed with the mechanical mode shape, until the gap between the disk and the center point of the cantilever reached a specific gap value. This was to simulate the real motion of the cantilever measured in the experiment. The deformation is kept small in order to stay in the linear regime. The resonant optical modes in the disk with the deformed cantilever were simulated by solving the eigenvalue problem of the optical field. After obtaining the resonant frequencies of a specific optical mode at different gaps, we find $g_{OM}$ as the slope of the fitted frequency-gap curve. We focus on the results obtained with 1st radial order optical modes, as these modes have the highest radiation-limited $Q_{opt}$ and are predicted to couple well to the cantilever mechanical modes of interest.

Alternatively, we can employ $g_{OM} = \omega_{opt}/L_{OM}$ and estimate the effective length $L_{OM}$ via the perturbative expression[28]

$$L_{OM} = \frac{2\int \varepsilon |\mathbf{E}|^2 dV}{\int (\mathbf{u}\cdot\mathbf{n})\left[\Delta\varepsilon |\mathbf{E}_\parallel|^2 - \Delta\varepsilon^{-1}|\mathbf{D}_\perp|^2\right]dA}, \qquad (4)$$

where, **E** and **D** are the modal electric and electric displacement fields respectively, $\Delta\varepsilon = \varepsilon_{diel} - \varepsilon_{air}$ and $\Delta\varepsilon^{-1} = (\varepsilon_{diel})^{-1} - (\varepsilon_{air})^{-1}$ where $\varepsilon_{diel}$ and $\varepsilon_{air}$ are the permittivities of the (dielectric) microdisk material and air, respectively. The mass displacement due to the mechanical resonance is given by **u**, and the normal surface displacement at the structure boundaries is $\mathbf{u} \cdot \mathbf{n}$, where **n** is the surface normal. In Eq. 4, **u** is normalized to the maximum displacement: $\mathbf{u} \rightarrow \mathbf{u}/|\mathbf{u}_{max}|$. In this design, The maximum displacement occurs at the tip of the cantilever. The integral in the denominator is performed over the entire surface of the nanostructure.

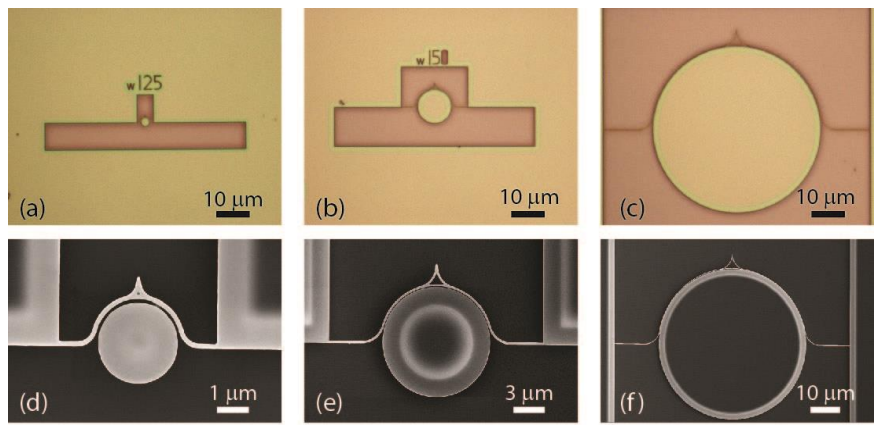

Fig. 5. Optical microscope images (a-c) and SEM images (d-f) of fabricated cantilever-disk devices. The disk diameter, $D$, and cantilever width, $w$, in the devices are: (a), (d) $D$ = 2.5 μm, $w$ = 132 nm ±6 nm; (b), (e) $D$ = 10 μm, $w$ = 172 nm ±5 nm; and (c), (f) $D$ = 50 μm, $w$ = 155 nm ±7 nm. Reprinted with permission from ref. 18.

## 1.4 Towards Optomechanical AFM Probes
### 1.4.1 Wide Range of Spring Constants

Commercial AFM cantilevers cover a wide range of spring constants $k$, from 0.01 N/m to over 1000 N/m. Cantilevers with soft spring constants usually find applications in imaging soft samples, such as biological materials, while the cantilevers with hard spring constants have proven useful in ultrahigh (atomic or even sub-atomic) resolution imaging. In Sec. 1.2.2, we have shown a cantilever in our integrated optomechanical system with a spring constant on the order of magnitude of 0.1 N/m. It is worthwhile investigating if the very soft and very hard regimes of typical spring constants are compatible with our disk-cantilever optomechanical configuration. By varying the microdisk diameter from 2.5 μm to 50 μm and the corresponding cantilever lengths accordingly (Fig. 5), we achieve spring constants ranging from 0.01 N/m to 300 N/m (Table 2). Meanwhile, the resonant frequency of the fundamental mode of such a cantilever is much higher than a conventional cantilever with a comparable spring constant. In particular, the fundamental mechanical frequencies (265 kHz to 111.4 MHz) are significantly higher than those of conventional AFM probes (10 kHz to 1.1 MHz)[30]. This high frequency range, enabled by the small mass of the nanoscale cantilevers, may increase the imaging acquisition rate, decrease thermal drift, reduce the impact of ambient vibration and acoustic noise, and enhance force sensitivity. More importantly, the displacement sensitivity is not greatly sacrificed for the cantilever with a very soft or very hard spring constant. The displacement sensitivity remains in the fm/Hz$^{1/2}$ range across the full range of cantilever stiffness. This is comparable to other state-of-the-art transduction schemes.

It is interesting to note that the device with $D$ = 2.5 μm has a low $Q_{opt}$ ($10^3$ to $10^4$) due to radiation losses; however, its optomechanical coupling $g_{OM}$ is much higher than the one with $D$ = 50 μm, because of a more confined mode volume (for $D$ = 2.5 μm). These two effects balance each other and the displacement sensitivities for $D$ = 2.5 μm and 50 μm are similar.

| $D$ (μm) | $w$ (nm) | $Q_{opt}$ | Exp. $f_m$ (MHz) | Cal. $f_m$ (MHz) | Cal. $m_{eff}$ (pg) | Cal. $k_{eff}$ (N/m) | Exp. $Q_m$ | Typical disp sens. (fm/(Hz)$^{1/2}$) ± 15% |
|---|---|---|---|---|---|---|---|---|
| 2.5 | 106±8 | 1.6×10$^4$ | 57.65 | 61 | 0.28 | 36 | 28 | |
|  | 132±6 | 7.2×10$^3$ | 68.78 | 75 | 0.34 | 64 | 66 | |
|  | 158±7 | 3.1×10$^3$ | 83.70 | 90 | 0.40 | 110 | 55 | 2.0 |
|  | 205±7 | 5.2×10$^3$ | 100.0 | 120 | 0.51 | 200 | 69 | |
|  | 238±12 | 1.2×10$^4$ | 111.4 | 140 | 0.60 | 290 | 80 | |
| 10 | 124±3 | 7.7×10$^4$ | 3.97 | 4.3 | 1.4 | 0.99 | 10 | |
|  | 149±3 | 3.7×10$^4$ | 4.77 | 5.1 | 1.7 | 1.5 | 13 | |
|  | 172±5 | 9.9×10$^4$ | 5.62 | 5.9 | 1.9 | 2.4 | 22 | 0.2 |
|  | 224±3 | 1.4×10$^4$ | 7.30 | 7.7 | 2.5 | 5.3 | 18 | |
|  | 256±4 | 8.7×10$^3$ | 8.17 | 8.8 | 2.9 | 7.6 | 21 | |
|  | 271±5 | 1.5×10$^5$ | 8.87 | 9.3 | 3.0 | 9.4 | 27 | |
| 50 | 107±5 | 1.6×10$^4$ | 0.265 | 0.27 | 3.8 | 0.011 | 1.1 | |
|  | 128±5 | 4.2×10$^4$ | 0.375 | 0.33 | 4.5 | 0.025 | 1.9 | |
|  | 155±7 | 3.4×10$^4$ | 0.433 | 0.40 | 5.5 | 0.041 | 1.7 | 1.0 |
|  | 210±5 | 3.4×10$^4$ | 0.538 | 0.54 | 7.4 | 0.085 | 2.5 | |
|  | 233±10 | 6.2×10$^4$ | 0.615 | 0.60 | 8.2 | 0.12 | 4.3 | |

Table 2. Experimentally measured and calculated properties of the disk-cantilever devices. The typical photodetector-limited displacement sensitivity numbers are taken for representative devices within each disk diameter range ($D$ = 2.5 μm, $w$ = 205 nm; $D$ = 10 μm, $w$ = 172 nm; and $D$ = 50 μm, $w$ = 155 nm). Reprinted with permission from ref. 18.

### 1.4.2 Towards Optomechanical AFM

The integrated silicon transducer we have discussed so far requires an external fiber taper waveguide and positioners to align the fiber taper in order to couple light into and out of the device. This makes it difficult to integrate with a scan system for AFM. Furthermore, the tip of the cantilever is in-plane and located far away from the edges of the substrate chip. Thus, it cannot be moved to the proximity of an off-chip surface under study. Moreover, the tip is not very sharp compared with conventional AFM cantilevers. In this section, we describe our recent development towards optomechanical AFM probes and show preliminary results where the optomechanical AFM probe works in contact mode.

The optomechanical AFM probe consists of a nanoscale cantilever, a microdisk, and on-chip waveguides (Fig. 6a). These elements are nanofabricated on an integrated chip. Batch fabrication at a wafer level is under development. Optical fibers are pigtailed, making external positioners for the fiber unnecessary. The tip of the cantilever is fabricated about 20 μm away from the edge of the substrate chip and protected by oxide. Using focused ion beam, the tip is exposed such that it was overhanging the chip edge, and we sharpened it using a focused ion beam before the final release by hydrofluoric acid. As displayed in Fig. 6b, the radius of the tip is found to be about 20 nm. The silicon microdisk acts as a whispering-gallery-mode optical cavity. The highest $Q_{opt}$ of the particular microdisk resonator ($D$ = 10 μm) in this experiment is observed to be about 57 000, which is probably limited by imperfect plasma etching in this particular sample. $Q_{opt}$ up to 1 million have been demonstrated for microdisks with similar designs[17,19]. As we discussed in Sec. 1.2, the motion of the cantilever changes the effective optical length and thereby modulates the resonant wavelength of the microdisk cavity. Fixing the laser wavelength at the shoulder of the resonance, we can measure the displacement of the cantilever based on the change in optical transmission intensity. The Brownian motion of the cantilever ($w$ = 100 nm and $G$ = 100 nm) is detected and the resonant frequency of the fundamental in-plane mode is about 3.5 MHz, with a typical quality factor of about 20.

Next, we demonstrate engagement of the probe to a sample surface under investigation. The sample (high-purity gold on mica) is mounted on a piezo scanner. The scanner sits on a stack of manual translational stages and slip-stick positioners. During the engagement, the laser wavelength is maintained on the shoulder of an optical resonance and the transmission intensity is monitored. Assuming we stay in the linear regime, the force between the tip and the surface is proportional to the change in the transmission intensity.

It is worth noting that this type of probe is more fragile than conventional cantilevers because of the nanoscale gap (≈ 100 nm) between the cantilever and the disk. If the approach procedure does not stop in time after the tip touches the surface, the cantilever will be translated too close to the microdisk and it may adhere to the sidewall of the microdisk even after the tip is separated from the surface. This phenomenon is known as stiction[31]. After stiction happens, we need to utilize a micro manipulator to separate the cantilever from the disk. This procedure is time-consuming and not always successful. In the experiment, in order to avoid this issue we use an in-situ scanning electron microscope (SEM) to perform the initial coarse positioning and employ the standard "woodpecker" approach in a slow and careful manner.

After the surface is located, we record force-distance curves (Fig. 6c). 'Snap-in' (marked by A) and 'pull-off' (marked by B) correspond to the tip jumping onto and off of the sample surface. Some hysteresis is observed, as is expected for the gold-on-mica sample. In order to image the topography of the sample, the position of the cantilever is then set to the repulsive-force regime (gray area in Fig. 6c). The optomechanical transducer works as an AFM probe in contact mode and the x/y piezo scanner is turned on to scan the surface. The measured displacement of the cantilever reflects nanoscale height changes on the surface. The topography image in Fig. 6d captures the slightly tilted nature of the gold-on-mica sample. The fluctuation in the image is mostly from the noise in the AFM setup rather than the surface roughness. Further investigation indicates that the mechanical noise from the home-made scanner system dominates the noise performance. Implementation of the optomechanical probes in commercial AFM systems is under way and is expected to suppress the mechanical noise from the scanner.

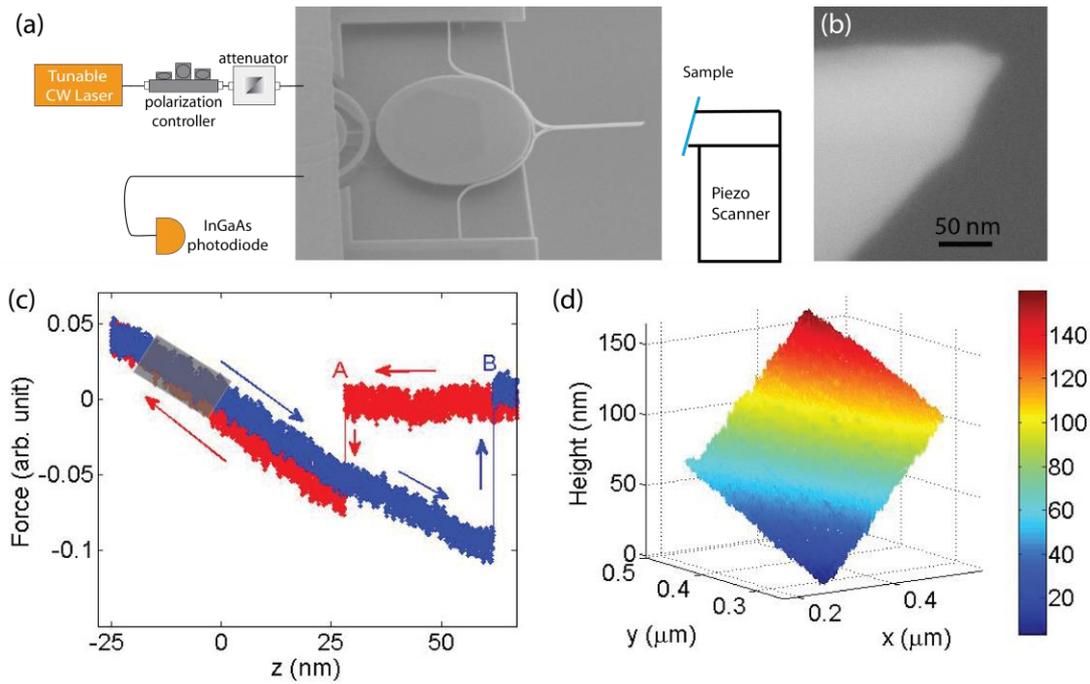

Fig. 6. (a) Schematics of the experimental setup. (b) Scanning electron micrograph of a sharpened tip. The radius is estimated to be about 20 nm. (c) A typical force-distance curve. 'Snap-in' (marked by A) and 'pull-off' (marked by B) occur when the tip jumps onto and off of the sample surface. The gray shadow indicates the regime where the tip is placed when operating in the contact mode. (d) The scan image of a slightly tilted flat surface of the gold-on-mica sample. © 2013 NSTI http://nsti.org. Reprinted and revised, with permission, from ref. 32.

## 1.5 Summary and Perspective

This chapter describes the basic design, simulation, and fabrication for fully Si integrated, waveguide coupled, optomechanical force and displacement sensors. The approach of full Si integration of all stationary nanophotonic components with mechanically separated movable components creates the opportunity to independently engineer these two parts for a variety of MEMS and NEMS sensing applications that require high precision, high bandwidth, and small footprint. Further integration of actuators for static and dynamic actuation is also possible.

The next development steps will be a system modification for scanning probe microscopy (SPM), which is a major application for mechanical motion detection. As the preferred method in scanning probe microscopy involves driving the probe near its mechanical resonance, it is necessary to an actuation mechanism for our probe. In most commercially available scanning probe microscopes, this is realized by an external driving piezoelectric element mounted near the cantilever[33]. The excitation takes place through mechanical stimulation by shaking the complete cantilever holder. This method is more difficult in liquid environments, because it generates standing waves through the acoustic stimulation of the liquid medium. Furthermore, it is challenging to excite narrow, high-frequency resonances in the MHz regime using this method. These drawbacks can be overcome by directly driving the cantilever with an integrated actuator. The most promising candidates to achieve an active oscillation control and full compatibility with commercially available scanning probe microscopes are electrostatic or thermal actuation, due to their relatively simple design integration and compatibility with the existing fabrication process.

We have already demonstrated the integration of an electrostatic actuator to tune the optical cavity resonance by 5.54 nm, which is useful for adjusting the device to operate with a fixed optical laser source, such as a compact and relatively inexpensive stabilized laser diode[19]. Using this electrostatic actuator, the optomechanical coupling and the readout sensitivity are tunable by two orders of magnitude for optimizing the sensor gain and dynamic range.

Alternatively, a thermal actuator can be realized with a bimorph structure, which is exposed to a temperature difference. Thermal actuation (also called bimorph actuation) requires two combined materials with a mismatch of the coefficient of thermal expansion (CTE). If heat is applied locally, the bimorph beam experiences a deflection. This can be achieved through a focused laser beam[34, 35] or Joule heating[36]. The advantages of bimorph actuation are the simplicity of implementation in the fabrication process and the lack of dependence on the substrate doping level.

Another interesting aspect of the device is the demonstrated cold damping of the mechanical degree of freedom by more than three orders of magnitude. Cold damping of the mechanical mode has been realized by applying an electrical feedback signal, which is derived from the optical readout signal, to an integrated electrostatic actuator. While not reducing the input-referred force noise, i.e. the Langevin force acting on the probe, the near-critical damping stabilizes the probe position and also flattens the frequency dependent sensor gain, allowing us to use the sensor effectively over a very broad frequency range without severe dynamic range constrains. The ability to dampen the mechanical noise can strongly reduce the backaction of the sensor onto a system being measured. Moreover, we achieved a noise level $\approx$2.3 times the standard quantum limit (SQL)[16, 37] for our mechanical system, approaching the fundamental readout limits. With future parameter improvement, it may be possible to cool the sensor to the quantum mechanical ground state while maintaining the high readout bandwidth[38] (i.e. in the 'bad' cavity limit).

It is worth noting the limitations of such a novel optomechanical probe. First, in order to pursue ultrahigh sensitivity, the dynamic range is sacrificed. Driving the cantilever at large amplitude, which is required in traditional tapping mode, will lead to nonlinear optomechanical response or mechanical stiction. It is necessary to fabricate different devices with much larger gaps between the cantilevers and the microdisks for this application. Secondly, the small gap may prevent the dynamic-mode operation of our probe in liquid environment due to high damping. Thirdly, the refractive index of the microdisk is temperature sensitive. Therefore, the thermal stability is required to achieve the high displacement sensitivity aforementioned.

In summary, we have demonstrated a novel class of fully-integrated cavity optomechanical transducers for mechanical position and motion sensing with high precision, high bandwidth and small footprint.

JZ, YL, and TM acknowledge support under the Cooperative Research Agreement between the University of Maryland and the National Institute of Standards and Technology Center for Nanoscale Science and Technology, Award 70NANB10H193, through the University of Maryland.